\documentclass[12pt]{article}
\newcommand{\beq}{\begin{equation}}
\newcommand{\eeq}{\end{equation}}
\newcommand{\beqa}{\begin{eqnarray}}
\newcommand{\eeqa}{\end{eqnarray}}
\newcommand{\beqar}{\begin{eqnarray*}}
\newcommand{\eeqar}{\end{eqnarray*}}
\newcommand{\hph}[1]{{\hphantom{#1}}}

\newcommand{\ga}{\gamma}
\newcommand{\Ga}{\Gamma}
\newcommand{\ka}{\kappa}
\newcommand{\inn}{\!\cdot\!}

\newcommand{\Om}{\Omega}

\newcommand{\z}{\zeta}

\newcommand{\eg}{{\it e.g.,}\ }
\newcommand{\ie}{{\it i.e.,}\ }
\newcommand{\labell}[1]{\label{#1}} 
\newcommand{\reef}[1]{(\ref{#1})}
\newcommand\prt{\partial}
\newcommand\veps{\varepsilon}

\newcommand\cR{{\cal R}}

\newcommand\cF{{\cal F}}

\newcommand\cM{{\cal M}}

\newcommand\cB{{\cal B}}

\newcommand\hR{\hat{R}}

\newcommand\tG{{\widetilde G}}

\newcommand\tC{{\widetilde C}}

\newcommand\Tr{{\rm Tr}}

\begin{document}

\begin{titlepage}

\begin{center}



\vskip 2 cm
{\LARGE \bf On   D-brane action at order $ \alpha'^2$  
 }\\
\vskip 1.25 cm
Ali Jalali\footnote{ali.jalali@stu-mail.um.ac.ir} and Mohammad R. Garousi\footnote{garousi@ um.ac.ir}

\vskip 1 cm
{{\it Department of Physics, Ferdowsi University of Mashhad\\}{\it P.O. Box 1436, Mashhad, Iran}\\}
\vskip .1 cm
\vskip .1 cm

\end{center}

\vskip 0.5 cm

\begin{abstract}
\baselineskip=18pt
We use compatibility of  D-brane  action  with linear T-duality, S-duality and with S-matrix elements as  guiding principles to find all world volume couplings of one massless closed    and two  open strings   at order  $\alpha'^2$ in type II superstring theories.
In particular, we find that the squares of   second fundamental form appear only in   world volume curvatures, and  confirm the observation that  dilaton   appears in  string frame action   via the transformation $\hat{R}_{\mu\nu}\rightarrow \hat{R}_{\mu\nu}+\nabla_{\mu}\nabla_{\nu}\Phi$. 
\end{abstract}

\end{titlepage}
\section{Introduction and Results}
The  low energy effective field theory of  D$_p$-branes  in type II superstring theories consists of the  Dirac-Born-Infeld (DBI) \cite{Bachas:1995kx} and the  Chern-Simons (CS) actions \cite{Douglas:1995bn}, \ie
\beqa
S_p&=&S_p^{DBI}+S_p^{CS}
\eeqa
 The curvature corrections to the DBI  action have been found in \cite{Bachas:1999um} by requiring consistency of the effective action with the $O(\alpha'^2)$ terms of the corresponding disk-level scattering amplitude \cite{Garousi:1996ad,Hashimoto:1996kf}. For totally-geodesic embedding of world-volume in  ambient spacetime in which  second fundamental form is zero, the corrections in  string frame  for zero B-field and for constant dilaton  are\footnote{Our index convention is that the Greek letters  $(\mu,\nu,\cdots)$ are  the indices of the space-time coordinates, the Latin letters $(a,d,c,\cdots)$ are the world-volume indices and the letters $(i,j,k,\cdots)$ are the normal bundle indices.}  
\beqa
S_p^{DBI} &\supset&-\frac{\pi^2\alpha'^2T_{p}}{48}\int d^{p+1}x\,e^{-\Phi}\sqrt{-\tG}\bigg[R_{abcd}R^{abcd}-2\hR_{ab}\hR^{ab}-R_{abij}R^{abij}+2\hR_{ij}\hR^{ij}\bigg]\labell{DBI}
\eeqa
where $\hR_{ab}=\tG^{cd}R_{cadb}$, $\hR_{ij}=\tG^{cd}R_{cidj}$ and $\tG=\det(\tG_{ab})$ where  $\tG_{ab}$  is the pull-back of  bulk metric onto the word-volume, \ie
\beqa
\tG_{ab}=\frac{\prt X^{\mu}}{\prt\sigma^a}\frac{\prt X^{\nu}}{\prt\sigma^b}G_{\mu\nu}\nonumber
\eeqa
The   Riemann curvatures in \reef{DBI} are the pull-back of the spacetime   curvature onto tangent and  normal bundles \cite{Bachas:1999um}.
 
The curvature corrections to the CS part can
be found by requiring that the chiral anomaly on the world volume of intersecting D-branes
(I-brane) cancels with the anomalous variation of the CS action \cite{Green:1996dd,Cheung:1997az,Minasian:1997mm}. These corrections  involve the  quadratic order of the curvatures at order $\alpha'^2$. However, the consistency of the effective action with the S-matrix elements of one NSNS and one RR vertex operators requires the CS part  at this order  to have   linear   curvature corrections as well \cite{Garousi:2010ki}, \ie
 \beqa 
S_p^{CS} &\supset&-\frac{\pi^2\alpha'^2T_{p}}{12}\int d^{p+1}x\, \epsilon^{a_0a_1\cdots a_p}\bigg[\frac{1}{(p+1)!}\nabla _j\cF^{(p+2)}_{ia_0\cdots a_p }\hat{R}^{ij}+\frac{1}{2!p!}\nabla_a\cF^{(p+2)}_{ija_1\cdots a_p }R_{a_0}{}^{aij}\bigg]\labell{CS}
\eeqa
where $\cF^{n+1}$ is the field strength of the RR potential $n$-form. The S-matrix calculations produce also the couplings in the CS part which involve linear field strength of B-field \cite{Garousi:2010ki} in which we are not interested in this paper.

For arbitrary embeddings,   the   couplings \reef{DBI} have been extended  in \cite{Bachas:1999um} to
\beqa 
S_p^{DBI} &\supset&-\frac{\pi^2\alpha'^2T_{p}}{48}\int d^{p+1}x\,e^{-\Phi}\sqrt{-\tG}\bigg[(R_T)_{abcd}(R_T)^{abcd}-2(\hR_T)_{ab}(\hR_T)^{ab}\nonumber\\
&&\qquad\qquad\qquad\qquad\qquad\qquad-(R_N)_{abij}(R_N)^{abij}+2\bar{R}_{ij}\bar{R}^{ij}\bigg]\labell{RTN}
\eeqa
where the world-volume curvature $(R_T)_{abcd}$ and $(R_N)^{abij}$   obey the Gauss-Codazzi equations, \ie
\beqa
(R_T)_{abcd}&=&R_{abcd}+\delta_{ij}(\Omega_{\ ac}{}^i\Omega_{\ bd}{}^j-\Omega_{\ ad}{}^i\Omega_{\ bc}{}^j)\nonumber\\
(R_N)_{ab}{}^{ ij}&=&R_{ab}{}^{ij}+g^{cd}(\Omega_{\ ac}{}^i\Omega_{\ bd}{}^j-\Omega_{\ ac}{}^j\Omega_{\ bd}{}^i)\labell{RTRN}
\eeqa
where $\Omega^i_{\ ab}$ is the second fundamental form \cite{Bachas:1999um}\footnote{Note that there is a minus sign typo on the right hand side of $(R_N)_{ab}{}^{ ij}$    in reference \cite{Bachas:1999um}. For totally-geodesic embedding, $(R_N)_{ab}{}^{ ij}$ must be equal $R_{ab}{}^{ij}$.} The relation between  $(\hR_T)_{ab}$ and the world volume curvature is then 
\beqa
(\hR_T)_{ab}=\hR_{ab}+\delta_{ij}(\Omega_c{}^c{}^i\Omega_{\ ab}{}^j-\Omega_{\ ca}{}^i\Omega_b{}^c{}^j)\labell{r1}
\eeqa
In equation \reef{RTN},  $\bar{R}^{ij}=\hR^{ij}+g^{ab}g^{cd}\Omega_{\ ac}{}^i\Omega_{\ bd}{}^j+\cdots$ where dots stand for unknown terms which involve the trace of the second fundamental form. They  could not be fixed in \cite{Bachas:1999um} because the couplings in \cite{Bachas:1999um} have been found by requiring the consistency of the corresponding couplings with the S-matrix element of one closed and two open string vertex operators for which the trace of the second fundamental form is zero. They may be fixed, however, by requiring the consistency of the couplings with dualities.
 
In static gauge and to the linear order of  fields, the second fundamental form has the following  simple form: 
\beqa
\Omega_{\ ab}{}^i=\prt_a\prt_b\chi^i +\Gamma^i_{ab} 
\eeqa
where $\chi^i$ is the massless transverse scalar field and $\Gamma^i_{ab} $ is the Levi-Civita connection.    The couplings of one graviton and two transverse scalars in \reef{RTN} have been shown    to be consistent with the corresponding  S-matrix elements \cite{Bachas:1999um}. However, there are couplings in \reef{RTN} which involve the trace of the second fundamental form which can not be checked with the S-matrix element of one closed and two open string vertex operators. We will show, among other things, that the trace term in  $(\hR_T)_{ab}$ is required by the consistency of the couplings \reef{RTN} with T-duality. Moreover,  we will find that the duality fixes the dots in $\bar{R}^{ij}$ to be 
\beqa
\bar{R}^{ij}=\hR^{ij}+g^{ab}g^{cd}(\Omega_{\ ac}{}^i\Omega_{\ bd}{}^j-\Omega_{\ ab}{}^i\Omega_{\ cd}{}^j)\labell{r2}
\eeqa
where the last term is the trace of the second fundamental form.

It has been observed in \cite{Garousi:2011fc,Garousi:2014oya} that the consistency of the closed string couplings with T-duality requires   the couplings of non-constant dilaton appear in  the world volume action via the transformation 
\beqa
\hR_{ab}\rightarrow \cR_{ab}=\hR_{ab}+\prt_a\prt_b\Phi&&\hR_{ij}\rightarrow\cR_{ij}=\hR_{ij}+\prt_i\prt_j\Phi\labell{crr}
\eeqa
We will find that the transformation of the couplings \reef{RTN} under the above replacement produces the couplings of one dilaton and two transverse scalars which are consistent with the dualities and with the corresponding S-matrix elements. In other worlds, the extension of the couplings \reef{DBI} to  include the curvature, the dilaton and the second fundamental form   are
\beqa 
S_p^{DBI} &\supset&-\frac{\pi^2\alpha'^2T_{p}}{48}\int d^{p+1}x\,e^{-\Phi}\sqrt{-\tG}\bigg[(R_T)_{abcd}(R_T)^{abcd}-2(\hat{\cR}_T)_{ab}(\hat{\cR}_T)^{ab}\nonumber\\
&&\qquad\qquad\qquad\qquad\qquad\qquad-(R_N)_{abij}(R_N)^{abij}+2\bar{\cR}_{ij}\bar{\cR}^{ij}\bigg]\labell{RTN1}
\eeqa
where $(\hat{\cR}_T)_{ab}$ and $\bar{\cR}_{ij} $ are the same as $(\hat{R}_T)_{ab}$ and $\bar{R}_{ij} $, respectively,  in which the replacement \reef{crr} have been performed.
We will show that similar extension exists for   the   couplings \reef{CS}, \ie the consistency of the couplings with dualities and with the S-matrix requires the following extension of \reef{CS}:
 \beqa 
S_p^{CS} &\!\!\!\supset\!\!\!&-\frac{\pi^2\alpha'^2T_{p}}{12}\int d^{p+1}x\, \epsilon^{a_0a_1\cdots a_p}\bigg[\frac{1}{(p+1)!}\nabla _j\cF^{(p+2)}_{ia_0\cdots a_p }\bar{\cR}^{ij}+\frac{1}{2!p!}\nabla_a\cF^{(p+2)}_{ija_1\cdots a_p }(R_N)_{a_0}{}^{aij}\bigg]\labell{CS1}
\eeqa
The coupling of the RR field strength and dilaton in the first term above has been already shown in \cite{Garousi:2011fc} to be consistent with the linear T-duality and with the S-matrix.   

  In general, one expects that the consistency of the world volume  couplings with full non-linear T-duality and S-duality would fix all   couplings at order $\alpha'^2$ \cite{Robbins:2014ara,Garousi:2014oya}, \eg the T-duality would relate the couplings \reef{CS1} to the standard CS couplings ${\cal{C}}^{p-3}(R_T\wedge R_T-R_N\wedge R_N)$ at order $\alpha'^2$. They would involve  also the world volume gauge field,  the spacetime  B-field and other RR-fields. In this paper, however, we will use only linear T-duality and S-duality. As a result, we will find many couplings which are consistent with such simplified dualities. We are interested in the couplings of one closed and two open string states in this paper. Even the coefficients of such couplings can not be fully fixed by the linear dualities. To reduce the number of arbitrary coefficients, we   use consistency of the couplings with the corresponding S-matrix elements as well. This latter condition fixes all unknown coefficients of the couplings in the DBI part, \ie we will find the couplings \reef{RTN1} and the following couplings in the string frame:
\beqa
S_p^{DBI} &\!\!\!\!\!\supset\!\!\!\!\!&-\frac{\pi^2\alpha'^2T_{p}}{12}\int d^{p+1}x\,e^{-\Phi}\sqrt{-\tG}\bigg[ \cR_{bd}\big(\prt_{a}F{^{ab}}\prt_{c}F^{cd}-\prt_{a}F_{c}{}^{d}\prt^{c}F^{ab}\big)+
\frac{1}{2}R_{bdce}\prt^{c}F^{ab}\prt^{e}F_{a}{}^{d}\nonumber\\
&& \qquad\qquad\qquad\qquad+\frac{1}{4}\cR_{d}{}^{d}\big(\prt_{a}F^{ab}\prt_{c}F_{b}{}^{c}+\prt_{b}F_a{}^{c}\prt_{c}F{}^{ab}\big)+\Om_{a}{}^{ai}\prt_{d}H_{c}{}^{d}{}{}_{i}\prt_{b}F^{bc}\nonumber\\
&& \qquad\qquad\qquad\qquad-\Om^{bai}\bigg(\prt_{b}F_{a}{}^{c}\prt_{d}H_{c}{}^{d}{}_{i}
+\prt^{d}F_{a}{}^{c}\prt_{i}H_{bcd}-\frac{1}{2}
\prt^{d}F_{a}{}^{c}\prt_{c}H_{bdi}\bigg)\bigg]\labell{DBI2}
\eeqa
where the scalar curvature $\cR_{a}{}^{a}\equiv \tG^{ab}\hR_{ab}+2\prt^a\prt_a\Phi$ is invariant under linear T-duality as the Ricci  curvatures $\cR_{ab}$ and $\cR_{ij}$ in \reef{crr}. The consistency of the couplings with the dualities and with the S-matrix elements fixes also the couplings in the CS part to be those in \reef{CS1} and the following couplings in the string frame:
\beqa
S_p^{CS} &\!\!\!\!\!\supset\!\!\!\!\!&\frac{\pi^2\alpha'^2T_{p}}{12}\int d^{p+1}x    \epsilon^{a_0a_1\cdots a_p}\bigg[ \frac{1}{2!(p-2)!}\prt^{a}F_{a_1a_2}\prt_{b}F_{a a_0}\prt^{b}\cF^{(p-2)}_{a_3a_4\cdots a_p}\labell{CS2}\\
&&\qquad\qquad\qquad-\frac{1}{(p-1)!} \Om_{a_0}{}^{ai} \prt_{a}F_{ba_1}\prt^{b}\cF^{(p)}_{ia_2a_3\cdots a_p} +\frac{1}{2!(p-1)!}\Om^{bai}\prt_{a}F_{a_0a_1}\prt_{b}\cF^{(p)}_{ia_2a_3\cdots a_p}\nonumber\\
&&\qquad\qquad\qquad-\frac{1}{2!(p-1)!}\Om_{a}{}^{ai} \prt^{b}F_{a_0a_1}\prt_{i}\cF^{(p)}_{ba_2a_3\cdots a_p}+\frac{1}{p!}\Om_{a}{}^{ai}\prt^{b}F_{ba_0}\prt_{i}\cF^{(p)}_{a_1a_2\cdots a_p}\nonumber\\
&&\qquad\qquad\qquad-\frac{1}{p!}\Om^{bai} \prt_{a}F_{ba_0}\prt_{i}\cF^{(p)}_{a_1a_2\cdots a_p}  +\frac{1}{(p-1)!} \Om_{a_0}{}^{ai}\prt^{b}F_{ba_1}\prt_{i}\cF^{(p)}_{aa_2a_3\cdots a_p}
\bigg]\nonumber
\eeqa
In the CS part, there is another multiplet   whose coefficient can not be fixed by the linear dualities and by the S-matrix elements of one closed and two open strings. It involves, however, the square of the second fundamental form.  On the other hand, as the couplings \reef{RTN1} and \reef{CS1} indicate, the square of the second fundamental form combines with the appropriate curvatures to form world volume curvatures $R_T$  and $\bar{R}$. Since the coefficients of    the curvature terms are already fixed in \reef{CS}, we expect the coefficient of this multiplet to be zero.

An outline of the paper is as follows: In the next section, we review the constraints that linear T-duality and S-duality may impose on an effective world volume action. In section 3,  we review the contact terms of the S-matrix element of one closed and two open strings at order $\alpha'^2$.    In section 4, we construct all coupling of one NSNS and two NS strings with arbitrary coefficients,  and find the coefficients by requiring the consistency of the couplings with the linear dualities and with the S-matrix elements.  In section 5, we construct all coupling of one RR and two NS strings with arbitrary coefficients,  and find the coefficients  by requiring the consistency of the couplings with the linear dualities and with the S-matrix elements.

\section{Linear duality constraints }

The T-duality and S-duality   transformations on massless field are in general nonlinear. Constraining the effective actions to be invariant under these nonlinear transformations which may fix all   couplings of bosonic fields including the non-perturbative effects \cite{Green:1997tv}, would be a difficult task ( see \cite{Liu:2013dna,Robbins:2014ara,Garousi:2014oya} for nonlinear T-duality). In this paper, however, we are interested only in the world volume couplings of one massless closed and two open string states at order $\alpha'^2$. Using the fact that the world volume couplings of one closed string and the couplings of one closed and one open strings have no higher derivative corrections in the superstring theory, one realizes that the higher derivative couplings of one   closed and two open string states   must be invariant under linear duality transformations.

The full set of nonlinear T-duality transformations has been found in \cite{TB,Meessen:1998qm,Bergshoeff:1995as,Bergshoeff:1996ui,Hassan:1999bv}. 
We consider a    background consists of    a constant dilaton $\phi_0$ and a metric which is  flat in all directions except the killing direction $y$ which is a circle with radius $\rho$. Assuming  quantum fields are small perturbations around this background, \eg $G_{\mu\nu}=\eta_{\mu\nu}+2h_{\mu\nu}$  and $G_{yy}=\frac{\rho^2}{\alpha'}(1+2h_{yy})$ where $\mu,\nu\neq y$, the T-duality transformations for the background are $e^{2\tilde{\phi_0}}=\frac{\alpha'e^{2 \phi_0}}{\rho^2}$,$ \tilde{G}_{\mu\nu}=\eta_{\mu\nu}$, $\tilde{G}_{yy}=\frac{\alpha'}{\rho^2}$ and the quantum fluctuations at the linear order take the following form\footnote{Note that if one considers full T-duality transformation for background and quantum fluctuations, then the effective action would contains all couplings at order $\alpha'^2$, \eg $H^4$ or $(\prt F)^2H^2$. However, in this paper we are interested  only in the couplings consisting of one closed   and two open string fields, hence we consider only   linear T-duality. }:
\beqa
&&\tilde{\phi}=\phi-\frac{1}{2}h_{yy},\,\tilde{h}_{yy}=-h_{yy},\, \tilde{h}_{\mu y}=B_{\mu y},\, \tilde{B}_{\mu y}=h_{\mu y},\,\tilde{h}_{\mu\nu}=h_{\mu\nu},\,\tilde{B}_{\mu\nu}=B_{\mu\nu}\nonumber\\
&&{  \tC}^{(n)}_{\mu\cdots \nu y}={  C}^{(n-1)}_{\mu\cdots \nu },\,\,\,{  \tC}^{(n)}_{\mu\cdots\nu}={  C}^{(n+1)}_{\mu\cdots\nu y}\labell{linear}
\eeqa
 The T-duality transformation of the world volume gauge field   when it is along the Killing direction, is $\tilde{A}_y=\chi_y$ where $\chi_y$ is the transverse scalar. Similarly,  $\tilde{\chi}_y=A_y$. When the gauge field and the transverse scalar field are not along the Killing direction, they are invariant under the T-duality.  We are interested in applying the above linear T-duality transformations on the quantum fluctuations and apply the full nonlinear T-duality on the    background. The latter requires the CS part to have no overall dilaton factor and the DBI part to have  the overall factor $e^{-\Phi}\sqrt{-\tilde{G}}$.

Following \cite{Garousi:2009dj}, the effective couplings which are invariant under the above linear T-duality can be constructed as follows: We first write,  in the static gauge, all couplings on the world volume of D$_p$-brane involving one massless closed and two open string states,   in terms of the world volume indices $a,b,\cdots$ and the transverse indices $i,j,\cdots$. We call this action $S_p$. Then we reduce the action to the 9-dimensional space. It produces two different actions. In one of them,  the Killing direction $y$ is a world volume direction, \ie $a=(\tilde{a},y)$,  which we call it $S_p^w$, and in the other one the Killing direction $y$ is a transverse direction, $i=(\tilde{i},y)$, which we call it $S_p^t$. The transformation of  $S_p^w$ under the linear T-duality \reef{linear} which we call it $S_{p-1}^{wT}$, must be equal to $S_{p-1}^t$ up to some total derivative terms, \ie
\beqa
S_{p-1}^{wT}-S_{p-1}^t&=&0\labell{Tconstraint}
\eeqa
This constrains the unknown coefficients in the original action $S_p$.

The S-duality of type IIB theory produces another set of constraints on the coefficients of $S_p$.  Under  the S-duality, the  graviton in  Einstein frame, \ie  $G^E_{\mu\nu}=e^{-\Phi/2}G_{\mu\nu}$, the transverse scalar fields  and the RR four-form are  invariant, and the following objects transform as doublets \cite{ Gibbons:1995ap,Tseytlin:1996it,Green:1996qg}:
\beqa\label{sdtrans}
\cB\ \equiv\ 
\pmatrix{B \cr 
C^{(2)}}\rightarrow (\Lambda^{-1})^T \pmatrix{B \cr 
C^{(2)}}\,\,\,\\
\cF\ \equiv\ \pmatrix{*F \cr 
G(F) }\rightarrow (\Lambda^{-1})^T \pmatrix{*F \cr 
G(F) }\nonumber
\eeqa
where the matrix $\Lambda \in SL(2,Z)$ and $G(F)$ is a nonlinear function of $F,\, \Phi,\, C$. To the linear order of the quantum fluctuations and nonlinear background which we call it linear S-duality\footnote{Note that we consider finite   $SL(2,Z)$ transformation but infinitesimal quantum fluctuations.},   $G(F)=e^{-\phi_0}F$ where $\phi_0$ is the constant dilaton background \cite{ Gibbons:1995ap}. In above equation   $(*F)_{ab}=\epsilon_{abcd}F^{cd}/2$.   The transformation of the dilaton and the RR scalar $C$ appears in the transformation of   the  $SL(2,Z)$ matrix $\cM$ 
\beqa
 {\cal M}=e^{\phi}\pmatrix{|\tau|^2\ C \cr 
C\ 1}\label{M}
\eeqa
where  $\tau=C+ie^{-\Phi}$. This matrix  transforms as \cite{ Gibbons:1995ap}
\beqa
{\cal M}\rightarrow \Lambda {\cal M}\Lambda ^T\labell{TM}
\eeqa
To the zeroth  and the first order of  quantum fluctuations and nonlinear order of the background field $\phi_0$, the matrix $\cM$ is
\beqa
 \cM_0=\pmatrix{e^{-\phi_0}& 0\cr 
0&e^{\phi_0}}\,\,,\,\,
\delta\cM=\pmatrix{-e^{-\phi_0}\phi& e^{\phi_0}C\cr 
 e^{\phi_0}C&e^{\phi_0} \phi}
\eeqa
They transform as \reef{TM} under the $SL(2,Z)$ transformations.

Using the above transformations, it is obvious that there must be no couplings in the Einstein frame between one dilaton and two transverse scalars because it is impossible to construct $SL(2,Z)$ invariant from $\cM_0$ and one $\delta\cM$, \ie $\Tr(\cM_0^{-1}\delta\cM)=0$. This produces a set of constraint on the coefficients of the effective action S$_p$.

One can easily found that the following structures are invariant under the linear S-duality transformation:
\beqa 
\prt (*\cF^T)\cM_0\prt^2\cB&=&e^{-\phi_0}\prt F\prt^2 B -\prt(*F) \prt^2C ^{(2)}  \nonumber\\
\prt\cF^T \cM_0\prt\cF &=& e^{-\phi_0}[\prt(*F)\prt(*F)+\prt F\prt F] \labell{doub}\\
\prt\cF^T\prt^2\cM\prt\cF&=& e^{-\phi_0}\prt^2\Phi\prt F\prt F-e^{-\phi_0}\prt^2\Phi\prt (*F) \prt (*F)
 +\prt^2 C\prt F\prt (*F)+\prt^2 C\prt (*F)\prt F\nonumber
\eeqa
Up to total derivative terms then the couplings of one closed and two open string states on the world volume of D$_3$-brane should appear in the structures $R\Omega\Omega$, $\prt^2 C^{(4)}\Omega\Omega$, $\Omega\prt(*\cF^T)\cM_0\prt^2\cB$, $R\prt\cF^T \cM_0\prt\cF $ and $\prt\cF^T\prt^2\cM\prt\cF$ which are invariant under the linear S-duality. They constrain the coefficients of the couplings in $S_p$.

\section{S-matrix constraints}
Another set of constraints on the coefficients of $S_p$ is produced by comparing the couplings with the S-matrix element of one closed and two open string states at order $\alpha'^2$. This S-matrix element has been calculated in \cite{Hashimoto:1996kf}
\beq
A\sim{\Ga[-2t]\over\Ga[1-t]^2}K(1,2,3)
\eeq
where $K$ is the kinematic factor   and $t=-\alpha' k_1\cdot k_2$ is the only Mandelstam variable in the amplitude. $k_1$ and $k_2$ are the   open string momenta.  The low energy expansion of the gamma functions is    ${\Ga[-2t]\over\Ga[1-t]^2}=-\frac{1}{2t}-\frac{\pi^2 t}{12}+\cdots$. The first term produces the couplings which are consistent with the corresponding couplings in DBI and CS actions at order $\alpha'^0$ \cite{Garousi:1998fg}. The second term produces the following on-shell   couplings   in the Einstein frame when the closed string is NSNS state \cite{Garousi:1998fg}: 
\beqa
A(\chi,\chi,h)&\!\!\!\!\sim\!\!\!\!&\nonumber\bigg(2k_1\cdot k_2\,\z_1\cdot\veps_3\cdot\z_2+k_1\cdot
k_2\,\z_1\cdot\z_2\,\veps_{3a}{}^a+\z_1\cdot p_3\,\z_2\cdot
p_3\,\veps_{3a}{}^a\\
&&\nonumber-2k_1\cdot\veps_3\cdot
k_2\,\z_1\cdot\z_2+4\z_1\cdot\veps_3\cdot k_1\,\z_2\cdot p_3
+(1\longleftrightarrow 2)\bigg)(k_1\inn k_2)^2 \\
\nonumber
A(\chi,\chi,\phi)&\!\!\!\!\sim\!\!\!\!&\frac{p-3}{2\sqrt{2}}\bigg(k_1\cdot
k_2\,\z_1\cdot\z_2+\z_1\cdot p_3\,\z_2\cdot p_3+(1\longleftrightarrow 2)\bigg)
(k_1\inn k_2)^2
\\
\nonumber
A(\chi,a,b)&\!\!\!\!\sim\!\!\!\!&-2i\bigg(2k_1^a\,\z_{1i}\,f_{2ab}\,\veps_3^{bi}-
\z_1\cdot p_3\,f_{2ab}\,\veps_3^{ab}\bigg)(k_1\inn k_2)^2\\
\nonumber
A(a,a,h)&\!\!\!\!\sim\!\!\!\!&2\bigg(\veps_{3ab}f_1{}^{ac}f_2{}^b{}_c
-{1\over4}f_{1ab}f_2{}^{ab}\,\veps_{3a}{}^a
+(1\longleftrightarrow 2)\bigg)(k_1\inn k_2)^2\\
\nonumber
A(a,a,\phi)&\!\!\!\!\sim\!\!\!\!&-\frac{p-7}{4\sqrt{2}}\bigg(f_{1ab}f_2{}^{ab}
+(1\longleftrightarrow 2)\bigg)(k_1\inn k_2)^2\labell{SDBI}
\eeqa
where $\z_1,\,\z_2$ are the polarizations of the open string states and $\veps_3$ is the polarization of the closed string. For the RR state, the couplings in the momentum space are \cite{Garousi:1998fg} 
\beqar
A(\chi,\chi,c_{(p+1)})
&\!\!\!\!\sim\!\!\!\!&-\frac{2}{(p+1)!}\bigg(\z_1\cdot p_{3}\,\z_2\cdot p_{3}\,\veps_{3}{}^{a_0...a_p}
+2(p+1)\,\z_1^i\,k_1^{a_0}\,\z_2\cdot p_3\,\veps_{3i}{}^{a_1...a_p}
\nonumber\\
&&\qquad\quad+
{p(p+1)}\,\z_1^i\,\z_2^j\,k_1^{a_0}\,k_2^{a_1}\,
\veps_{3ij}{}^{a_2...a_p}\bigg)\,\epsilon^v_{a_0...a_p}
(k_1\inn k_2)^2+(1\longleftrightarrow 2)\nonumber\\
A(\chi,a,c_{(p-1)})&\!\!\!\!\sim\!\!\!\!&-\frac{2}{(p-1)!}\bigg(\z_1\cdot p_3\,f_2^{a_0a_1}\,
\veps_3{}^{a_2...a_p}+(p-1)\z_1^i\,f_2^{a_0a_1}\,k_1^{a_2}\,
\veps_{3i}{}^{a_3...a_p}\bigg)\,\epsilon^v_{a_0...a_p}
(k_1\inn k_2)^2\nonumber\\
A(a,a,c_{(p-3)})&\!\!\!\!\sim\!\!\!\!&-\frac{1}{2(p-3)!}f_1{}^{a_0a_1}f_2{}^{a_2a_3}
\veps_3{}^{a_4...a_p}\,\epsilon^v_{a_0...a_p}
(k_1\inn k_2)^2+(1\longleftrightarrow 2)\labell{SCS}
\eeqar
Compatibility of the  couplings  with above   amplitudes constrains the coefficients in $S_p$.

It has been   argued in \cite{Robbins:2014ara} that to construct the effective action for probe branes, one has to impose   the bulk equations of motion  at order $\alpha'^0$ into $S_p$. Since we are interested in the world volume couplings which have linear closed string fields, we have to impose the supergravity equations of motion at linear order, \ie
\beqa
R+4\nabla^2\Phi&=&0\nonumber\\
R_{\mu\nu}+2\nabla_{\mu\nu}\Phi&=&0\nonumber\\
\nabla^{\rho}H_{\rho\mu\nu}&=&0\nonumber\\
\nabla^{\mu_1}\cF^{(n)}_{\mu_1\mu_2\cdots \mu_n}&=&0
\eeqa
where $\mu,\nu,\rho$ are the bulk indices. Using these equations, one finds
\beqa
R_{\mu\hph{i}\nu i}^{\hph{a}i} &=&  -2\nabla_{\mu\nu}\Phi-R_{\mu\hph{c}\nu c}^{c}\nonumber\\
\nabla^i{}_i\Phi&=&-\nabla^a{}_a\Phi\nonumber\\
\nabla^{i}H_{i\mu\nu}&=&-\nabla^{a}H_{a\mu\nu}\nonumber\\
\nabla^{i}\cF^{(n)}_{i\mu_2\cdots \mu_n}&=&-\nabla^{a}\cF^{(n)}_{a\mu_2\cdots \mu_n}\labell{eom}
\eeqa
which indicates that the terms on the left-hand side are not    independent. In other words, the coefficients of the couplings in $S_p$ which involve the terms on the left-hand side above must be zero. 

\section{DBI couplings}

In this section, using the mathematica package ``xAct'' \cite{CS}, we are going to write all couplings of one closed string NSNS state and two open strings   with unknown coefficients. We then constrain the coefficients by imposing the consistency of the couplings with the linear dualities and with the corresponding S-matrix element.  Since all such couplings are too many to be written them once, we consider the couplings with specific closed string NSNS state and open string NS states. 

\subsection{One graviton and two  transverse scalar fields}

We begin with the couplings of one graviton and two transverse scalar fields. The transverse scalar fields should appear in the action   through the pull-back of bulk tensors, through the Taylor expansion of bulk tensors or through  the second fundamental form. Since there is no  higher-derivative correction to the  couplings of  one closed string and one open string in type II superstring theories, \eg there is no coupling with structure $DR\Om$ or $RD\Om$, the pull-back operator and Taylor expansion would produce no coupling between two scalars and one curvature from $DR\Om$ or $RD\Om$. Therefore, the only possibility for the two transverse scalars is through the second fundamental form.   All such   couplings at order $\alpha'^2$ are  the following:
\beqa 
S_{h\chi\chi} &\!\!\!\!\!\!=\!\!\!\!\!\!&\frac{\pi^2\alpha'^2T_{p}}{12}\int d^{p+1}x\,e^{-\Phi}\sqrt{-\tG}\bigg[w_1 R^{bc}{}_{bc} \Omega_{a}{}^{ai} \Omega_{d}{}^{d}{}_{i}+ 2 w_2 R^{bj}{}_{bj} \Omega_{a}{}^{ai} \Omega_{c}{}^{c}{}_{i} +w_3 R^{ij}{}_{ij} \Omega_{a}{}^{ak} \Omega_{b}{}^{b}{}_{k} \nonumber\\
&&\qquad\qquad\qquad\qquad\qquad\quad+w_4 R^{b}{}_{ibj} \Omega_{a}{}^{ai} \Omega_{c}{}^{cj} + w_5 R_{i}{}^{j}{}_{kj} \Omega_{a}{}^{ai} \Omega_{b}{}^{bk}+ w_6 R^{bc}{}_{bc} \Omega_{dai} \Omega^{dai}   \nonumber\\
&&\qquad\qquad\qquad\qquad\qquad\quad+2 w_7 R^{bj}{}_{bj} \Omega_{cai} \Omega^{cai} + w_8 R^{kj}{}_{kj} \Omega_{bai} \Omega^{bai}+ w_9 R^{b}{}_{ibj} \Omega_{ca}{}^{j} \Omega^{cai}  \nonumber\\
&&\qquad\qquad\qquad\qquad\qquad\quad+w_{10} R_{i}{}^{j}{}_{kj} \Omega_{ba}{}^{k} \Omega^{bai} + w_{11} R_{d}{}^{c}{}_{bc} \Omega^{b}{}_{ai} \Omega^{dai} +w_{12} R_{c}{}^{j}{}_{bj} \Omega^{b}{}_{ai} \Omega^{cai}  \nonumber\\
&&\qquad\qquad\qquad\qquad\qquad\quad+w_{13} R_{cbij} \Omega^{b}{}_{a}{}^{j} \Omega^{cai}+ w_{14} R_{cibj} \Omega^{b}{}_{a}{}^{j} \Omega^{cai}-w_{15} R_{cjbi} \Omega^{b}{}_{a}{}^{j} \Omega^{cai}  \nonumber\\
&&\qquad\qquad\qquad\qquad\qquad\quad+w_{16} R_{d}{}^{c}{}_{bc} \Omega_{a}{}^{ai} \Omega^{bd}{}_{i}+ w_{17} R_{c}{}^{j}{}_{bj} \Omega_{a}{}^{ai} \Omega^{bc}{}_{i} + w_{18} R_{cibj} \Omega_{a}{}^{ai} \Omega^{bcj}  \nonumber\\
&&\qquad\qquad\qquad\qquad\qquad\quad+w_{19} R_{abdc} \Omega^{cb}{}_{i} \Omega^{dai}\bigg]\labell{hpp}
\eeqa
Where $w_i$ with $ i=1,2,\cdots,19$ are the unknown constants that must be determined by imposing various constraints.

All above couplings are not  independent. In fact by applying the cyclic symmetry of the Riemann curvature, one can neglect  some of the constants.  
For example, one finds  
 the coupling in \reef{hpp} with coefficient $w_{13}$, $w_{14}$ and $w_{15}$ are not independent, \ie
\beqa 
&&w_{13} R_{cbij} \Omega^{b}{}_{a}{}^{j} \Omega^{cai}+ w_{14} R_{cibj} \Omega^{b}{}_{a}{}^{j} \Omega^{cai}-w_{15} R_{cjbi} \Omega^{b}{}_{a}{}^{j} \Omega^{cai}=\nonumber\\
&&(w_{13}+w_{15}) R_{cbij} \Omega^{b}{}_{a}{}^{j} \Omega^{cai}+  (w_{14}-w_{15}) R_{cibj} \Omega^{b}{}_{a}{}^{j} \Omega^{cai}  
\eeqa
So the coupling with coefficient   $w_{15}$ is not independent and may be ignored from the list \reef{hpp} before imposing various  constraints. Alternatively, one may keep all couplings in \reef{hpp} and imposes the constraints to find appropriate relations between the coefficients   and at the end imposes the cyclic symmetry. The final result of course must be identical in both methods. However, we find the latter method is easier to apply by computer so we do it in this paper. In fact after imposing the constraints, we write the     Riemann curvature in terms of metric. Then all terms that are related by the cyclic symmetry would be canceled. So the coefficients of all such terms can easily be set to zero.

By comparing the above couplings with \reef{RTN} we find $w_9=1, w_{11}=1, w_{16}=-1$ and $w_{19}=1$. These constraints are in fact the S-matrix constraints because the couplings in \reef{RTN} are fixed in \cite{Bachas:1999um} by  comparing them with the corresponding S-matrix elements.  Furthermore, the constraint that the bulk equations of motion \reef{eom} have to be imposed on the brane couplings, fixes the coefficients
$w_2=w_3=w_5=w_7=w_8=w_{10}=w_{12}=w_{17}=0$.

\subsection{One graviton and two gauge fields}

Under T-duality, the transverse scalar field along  the Killing direction transforms to the gauge field, \ie $\Om$ transform to $\prt F$. So  consistency of the couplings \reef{hpp} with T-duality requires the couplings of one graviton and two gauge fields to have structure $R\prt F\prt F$.   All   such couplings    are  the following: 
\beqa
&&S_{haa} =\frac{\pi^2\alpha'^2T_{p}}{12}\int d^{p+1}x\,e^{-\Phi}\sqrt{-\tG}\bigg[z_1 R^{cd}{}_{cd} \partial_{a}F^{ae} \partial_{b}F_{e}{}^{b}+ 2 z_2 R^{ci}{}_{ci} \partial_{a}F^{ad} \partial_{b}F_{d}{}^{b}\labell{haa}\\
&&+ z_3 R^{ij}{}_{ij} \partial_{a}F^{ac} \partial_{b}F_{c}{}^{b}+ z_4 R_{e}{}^{d}{}_{cd} \partial_{a}F^{ae} \partial_{b}F^{bc}+z_5 R_{d}{}^{i}{}_{ci} \partial_{a}F^{ad} \partial_{b}F^{bc}+ z_6 R_{e}{}^{d}{}_{cd} \partial_{a}F_{b}{}^{c} \partial^{b}F^{ae}\nonumber\\
&&+ z_7 R_{d}{}^{i}{}_{ci} \partial_{a}F_{b}{}^{c} \partial^{b}F^{ad}+z_8 R^{cd}{}_{cd} \partial^{b}F^{ae} \partial_{e}F_{ab}+2 z_9 R^{ci}{}_{ci} \partial^{b}F^{ad} \partial_{d}F_{ab}+z_{10} R^{ij}{}_{ij} \partial^{b}F^{ac} \partial_{c}F_{ab}\nonumber\\
&&+z_{11} R^{cd}{}_{cd} \partial_{b}F_{ae} \partial^{b}F^{ae}+2 z_{12} R^{ci}{}_{ci} \partial_{b}F_{ad} \partial^{b}F^{ad}+z_{13} R^{ij}{}_{ij} \partial_{b}F_{ac} \partial^{b}F^{ac}+ z_{14} R_{e}{}^{d}{}_{cd} \partial_{b}F_{a}{}^{c} \partial^{b}F^{ae}\nonumber\\
&&+z_{15} R_{d}{}^{i}{}_{ci} \partial_{b}F_{a}{}^{c} \partial^{b}F^{ad}+z_{16} R_{aecd} \partial_{b}F^{cd} \partial^{b}F^{ae}+z_{17} R_{aced} \partial_{b}F^{cd} \partial^{b}F^{ae}+z_{18} R_{b}{}^{d}{}_{cd} \partial^{b}F^{ae} \partial^{c}F_{ae}\nonumber\\
&&+z_{19} R_{b}{}^{i}{}_{ci} \partial^{b}F^{ad} \partial^{c}F_{ad}+z_{20} R_{e}{}^{d}{}_{cd} \partial^{b}F^{ae} \partial^{c}F_{ab}+z_{21}R_{d}{}^{i}{}_{ci} \partial^{b}F^{ad} \partial^{c}F_{ab}+z_{22} R_{b}{}^{d}{}_{cd} \partial_{a}F^{ae} \partial^{c}F_{e}{}^{b}\nonumber\\
&&+z_{23} R_{b}{}^{i}{}_{ci} \partial_{a}F^{ad} \partial^{c}F_{d}{}^{b}+z_{24}R_{ebcd} \partial^{b}F^{ae} \partial^{d}F_{a}{}^{c}+z_{25}R_{ecbd} \partial^{b}F^{ae} \partial^{d}F_{a}{}^{c}+z_{26}R_{edbc} \partial^{b}F^{ae} \partial^{d}F_{a}{}^{c}\nonumber\\
&&+z_{27} R_{aecd} \partial^{b}F^{ae} \partial^{d}F_{b}{}^{c}+z_{28}R_{aced} \partial^{b}F^{ae} \partial^{d}F_{b}{}^{c}+z_{29}R_{ebcd} \partial_{a}F^{ae} \partial^{d}F^{bc}+z_{30}R_{edbc} \partial_{a}F^{ae} \partial^{d}F^{bc}  \bigg]\nonumber
\eeqa
Where $z_i$ with $ i=1,2,\cdots,30$ are constants   that must be determined by imposing the constraints  and $F^{ab}$ is field strength of the gauge field. Here also one may impose the cyclic symmetry and the Bianchi identity $dF=0$ before  imposing the constraints to cancel some of the couplings in \reef{haa} before . However, we prefer to impose  the cyclic symmetry and the bianchi identity after imposing the constraints. The bulk equations of motion \reef{eom} constrain   $z_2= z_3=z_5= z_7= z_9= z_{10}= z_{12}= z_{13}= z_{15}= z_{19}= z_{21}= z_{23}=0$.
 
\subsection{One dilaton and two transverse scalar fields}

The same reason as in section 4.1, leads one to conclude that the couplings of one dilaton and two transverse scalar fields have  structure $\prt\prt\Phi\Om\Om$. All   such  couplings   are  the following: 
\beqa 
S_{\Phi\chi\chi}&=&\frac{\pi^2\alpha'^2T_{p}}{12}\int d^{p+1}x\,e^{-\Phi}\sqrt{-\tG}\bigg[t_1 \Omega^{a}{}_{a}{}^{i}\Omega^{b}{}_{bi} \partial_{c}\partial^{c}\Phi+t_2 \Omega^{a}{}_{a}{}^{i} \Omega^{bc}{}_{i}\partial_{c}\partial_{b}\Phi+t_3 \Omega_{a}{}^{c}{}_{i} \Omega^{abi} \partial_{c}\partial_{b}\Phi\nonumber\\
&&\qquad\qquad\qquad\qquad\qquad\quad+t_4 \Omega_{abi} \Omega^{abi}\partial_{c}\partial^{c}\Phi+ t_5 \Omega^{a}{}_{a}{}^{i}\Omega^{b}{}_{bi} \partial_{j}\partial^{j}\Phi+ t_6 \Omega_{abi} \Omega^{abi}\partial_{j}\partial^{j}\Phi\nonumber\\
&&\qquad\qquad\qquad\qquad\qquad\quad+t_7\Omega^{a}{}_{a}{}^{i} \Omega^{b}{}_{b}{}^{j}\partial_{j}\partial_{i}\Phi +t_8\Omega_{ab}{}^{j} \Omega^{abi} \partial_{j}\partial_{i}\Phi \bigg]\labell{Ppp}
\eeqa
Where $t_i$ with $ i=1,2,\cdots,8$ are the unknown constants that we must be determined. The bulk equations of motion  \reef{eom} constrain $t_5=t_6=0$.

\subsection{One dilaton and two gauge fields }

  The  consistency of the couplings \reef{Ppp} with T-duality requires the couplings of one dilaton   and two gauge fields to have structure $\prt\prt\Phi\prt F\prt F$.   All   such couplings    are  the following: 
\beqa
S_{\Phi aa}&\!\!\!\!\!=\!\!\!\!\!&\frac{\pi^2\alpha'^2T_{p}}{12}\int d^{p+1}x\,e^{-\Phi}\sqrt{-\tG}\bigg[x_1\partial_{a}F^{cd} \partial_{b}F_{cd}\partial^{b}\partial^{a}\Phi + x_2 \partial^{b}\partial^{a}\Phi\partial_{c}F_{a}{}^{c} \partial_{d}F_{b}{}^{d}\nonumber\\&&\qquad\qquad\qquad + x_3\partial_{a}\partial^{a}\Phi \partial_{b}F^{bc}\partial_{d}F_{c}{}^{d} 
+ x_4 \partial_{b}F_{a}{}^{c}\partial^{b}\partial^{a}\Phi \partial_{d}F_{c}{}^{d} + x_5\partial_{b}F_{cd} \partial^{b}\partial^{a}\Phi \
\partial^{d}F_{a}{}^{c} \nonumber\\&&\qquad\qquad\qquad+ x_6 \partial^{b}\partial^{a}\Phi\partial_{c}F_{bd} \partial^{d}F_{a}{}^{c}
 + x_7\partial^{b}\partial^{a}\Phi \partial_{d}F_{bc}\partial^{d}F_{a}{}^{c} + x_8 \partial_{a}\partial^{a}\Phi\partial_{c}F_{bd} \partial^{d}F^{bc}\nonumber\\&&\qquad\qquad\qquad + x_9 \partial_{a}\partial^{a}\Phi \partial_{d}F_{bc} \partial^{d}F^{bc} 
 + x_{10} \partial_{a}F^{ab}\partial_{c}F_{b}{}^{c} \partial_{i}\partial^{i}\Phi + x_{11}\partial_{b}F_{ac} \partial^{c}F^{ab} \partial_{i}\partial^{i}\Phi \nonumber\\&&\qquad\qquad\qquad+x_{12} \partial_{c}F_{ab} \partial^{c}F^{ab} \partial_{i}\partial^{i}\Phi\bigg]\labell{Paa}
 \eeqa
where the constants $x_i$ with $\ i=1,2,\cdots,12$ must be determined by imposing the constraints.  The bulk equations of motion  \reef{eom} constrain $x_{10}=x_{11}=x_{12}=0$.

\subsection{One B-field, one transverse scalar field and one gauge field }

The final list of couplings in the DBI part is the  couplings of one B-field, one transverse scalar field and one gauge field which is the following: 
\beqa\label{bpf}
S_{ba\chi}&\!\!\!\!\!=\!\!\!\!\!&\frac{\pi^2\alpha'^2T_{p}}{12}\int d^{p+1}x\,e^{-\Phi}\sqrt{-\tG}\bigg[\gamma_1\Omega^{abi}\partial_{a}F^{cd}\partial_{b}H_{cdi}+\gamma_2\Omega^{abi}\partial_{b}H_{adi}\partial_{c}F^{cd}\labell{bap}\\
&&\qquad\qquad\qquad\qquad+\gamma_3\Omega^{abi}\partial_{a}F^{cd}\partial_{d}H_{bci}+\gamma_4\Omega^{abi}\partial_{c}F_{a}{}^{c}\partial_{d}H_{b}{}^{d}{}_{i}+\gamma_6\Omega^{abi}\partial_{b}F_{a}{}^{c}\partial_{d}H_{c}{}^{d}{}_{i}\nonumber\\
&&\qquad\qquad\qquad\qquad+\gamma_5\Omega^{a}{}_{a}{}^{i}\
\partial_{b}F^{bc}\partial_{d}H_{c}{}^{d}{}_{i}+\gamma_7\Omega^{abi}\partial_{b}H_{cdi}\partial^{d}F_{a}{}^{c}+\gamma_8\Omega^{abi}\partial_{c}H_{bdi}\partial^{d}F_{a}{}^{c}\nonumber\\
&&\qquad\qquad\qquad\qquad+\gamma_9\Omega^{abi}\partial_{d}H_{bci}\partial^{d}F_{a}{}^{c}+\gamma_{10}\Omega^{a}{}_{a}{}^{i}\partial_{c}H_{bdi}\partial^{d}F^{bc}+\gamma_{11}\Omega^{a}{}_{a}{}^{i}\partial_{d}H_{bci}\partial^{d}F^{bc}\nonumber\\
&&\qquad\qquad\qquad\qquad+\gamma_{12}\Omega^{abi}\partial_{a}F^{cd}\partial_{i}H_{bcd}+\gamma_{13}\Omega^{abi}\partial^{d}F_{a}{}^{c}\partial_{i}H_{bcd}+\gamma_{14}\Omega^{a}{}_{a}{}^{i}\partial^{d}F^{bc}\partial_{i}H_{bcd}\nonumber\\
&&\qquad\qquad\qquad\qquad-\gamma_{15}\Omega^{abi}\partial_{c}F_{a}{}^{c}\partial_{j}H_{bi}{}^{j}-\gamma_{17}\Omega^{abi}\partial_{b}F_{a}{}^{c}\partial_{j}H_{ci}{}^{j}-\gamma_{16}\Omega^{a}{}_{a}{}^{i}\partial_{b}F^{bc}\partial_{j}H_{ci}{}^{j}\bigg]\nonumber
\eeqa
Where $\gamma_i$ with $ i=1,2,\cdots,17$ are the unknown constants. The equations of motion \reef{eom} fixes  $\ga_{15}=\ga_{16}=\ga_{17}=0$.

We now consider the sum of the couplings in \reef{hpp}, \reef{haa}, \reef{Ppp}, \reef{Paa} and \reef{bap}, \ie
\beqa
S^{DBI}_p&=&S_{h\chi\chi}+S_{haa}+S_{\Phi\chi\chi}+S_{\Phi aa}+S_{ba\chi}\labell{tot1}
\eeqa
and apply the T-duality constraint \reef{Tconstraint}. It gives the following relations between the  constants:
\beqa
&&t_8=1,\ t_3=-t_2,\ t_1=-\frac{t_2}{2}-\frac{x_4}{2}-x_3,\ t_4=\frac{t_2}{2}+\frac{x_4}{2}+x_8-2x_9\nonumber\\
&& t_7=-1-z_{14}-z_4-z_6,\ w_1=\frac{1}{4}-\frac{x_3}{2}-\frac{x_4}{4},\ w_{15}=2+w_{14}-2\ga_1+\ga_3+2\ga_6+\ga_7-\ga_9\nonumber\\
&& w_{18}=2+2z_{14}-z_{29}+2z_{30}+2z_6,\ w_4=-1-z_{14}-z_4-z_6,\ x_5=2x_1-x_4+x_7\nonumber\\
&&z_{20}=-z_{14}-2z_{18}+z_{22},\ \ga_{13}=w_{14}+2\ga_{12}+\ga_3+\ga_6-\ga_9,\ \ga_5=z_{14}+z_4+z_6-\ga_6\nonumber\\
&&z_{28}=2z_{14}+4z_{16}+2z_{17}+z_{24}-z_{26}-2z_{27}+2z_6+2\ga_1-\ga_3-2\ga_6-\ga_7+\ga_9\nonumber\\ 
&&\ga_8=\frac{1}{2}-\ga_4-\ga_9,\ w_6=-\frac{1}{4}+\frac{x_4}{4}+\frac{x_8}{2}+x_9,\ z_1=\frac{x_3}{2}+\frac{x_4}{4}-\frac{z_{22}}{4},\ z_{25}=\frac{1}{2}+\frac{z_{22}}{2}-z_{26}\nonumber\\
&&z_{8}=\frac{x_4}{4}+\frac{x_8}{2}-\frac{z_{22}}{4}+x_9-2z_{11},\ \ga_2=-\frac{z_{29}}{2}+z_{14}+z_{30}+z_6-\ga_6\nonumber\\
&&\ga_{11}=\frac{z_{14}}{2}-\frac{z_{29}}{4}+\frac{z_{30}}{2}+\frac{z_{6}}{2}-\frac{\ga_{10}}{2}-\frac{\ga_{6}}{2},\ x_6=-x_2-x_7+z_{14}+z_4+z_6
\eeqa
As can be seen, not all coefficients of the DBI part are fixed by imposing consistency of the couplings with the linear T-duality, so we need further constraints which may be the consistency with S-duality. 

 In general, S-duality connect the DBI couplings containing the NSNS states to the CS couplings containing RR states. However, the S-duality constrains even the couplings in the DBI part. For example, the world volume couplings of D$_3$-brane in the Einstein frame must have no coupling with structure $\Phi\Om\Om$. This produces the following constraints: 
\beqa
&&\ga_9=2-2\ga_1+\ga_3+2\ga_6+\ga_7,\ x_9=\frac{1}{4}-\frac{x_4}{4}-\frac{x_8}{2}\\\nonumber
&& z_4=\frac{1}{2}+x_3+\frac{x_4}{2}-\frac{z_{29}}{2}+z_{30},\ z_6=-1-z_{14}-z_{30}+\frac{z_{29}}{2}
\eeqa
Another constraint   from the S-duality in the DBI part is that up to total derivative terms, the couplings of one graviton and two gauge fields in  D$_3$-brane action must appear in the S-duality invariant structure   $R\prt\cF^T\cM_0\prt\cF=e^{-\phi_0}R(\prt (*F)\prt (*F)+\prt F \prt F )$. This produces the following constraints: 
\beqa
z_{14}=0,\ x_4=1-2x_3
\eeqa
The S-duality constrains  the couplings of one dilaton and two gauge fields. It also   connects them to the couplings of one RR scalar and two gauge fields. This is resulted from the fact that the S-duality invariant structure which contains  the couplings of one dilaton and two gauge fields is $\prt\cF^T\prt^2\cM\prt\cF=e^{-\phi_0}\prt^2\Phi(-\prt (*F)\prt (*F)+\prt F \prt F )+\cdots$ where dots refer to the RR scalar couplings. This constraint on the couplings of one dilaton and two gauge fields produces the following relation: 
\beqar
x_8=x_3
\eeqar
The S-duality connects the DBI couplings of one B-field, one gauge field and one transverse scalar field to the CS couplings of one RR two-form, one gauge field and one transverse scalar. In the next section we will write all couplings in the CS part and impose the T-duality condition \reef{Tconstraint}. Then we will impose the above S-duality   condition. It produces  the following relation between the coefficients in the DBI part: 
\beqa
\ga_6=-1+2\ga_1-\ga_3\labell{last}
\eeqa
and many relations between the coefficients in the CS part (see constraints in \reef{SCS1}).

Imposing the above relations between the coefficients in $S^{DBI}_p$, we find that the action \reef{tot1}  are consistent with the S-matrix elements in \reef{SDBI} except the following terms:
\beqa
w_{14}\bigg(R_{bicj}\Om^{bai}\Om^{c}{}_{a}{}^{j}-R_{bjci}\Om^{bai}\Om^{c}{}_{a}{}^{j}+\Om^{bai}\prt^{d}F_{a}{}^{c}\prt_{i}H_{bcd}\bigg)
\eeqa
They are not consistent with the couplings in \reef{RTN} and with the corresponding S-matrix elements, so 
\beqa
w_{14}=0
\eeqa
As can be seen, there are still many coefficients which are not fixed by the linear dualities and with the S-matrix elements. 

 We have considered all couplings in $S^{DBI}_p$ which contains the Riemann curvature and the first derivative of the field strengths of the gauge field and the B-field. The  Riemann curvature satisfies the cyclic symmetry and the field strengths satisfy the Bianchi identities. So we have to impose these symmetries in   $S^{DBI}_p$. To perform this step, we write all field strengths in  terms of their corresponding  potentials and write the Riemann curvature in terms of 
\beqa
R_{abcd}=\prt_b\prt_ch_{ad}+\prt_a\prt_d h_{bc}-\prt_b\prt_d h_{ac}-\prt_a\prt_c h_{bd}
\eeqa
Then we find the coefficients    $\ga_1,\ \ga_3,\ \ga_7,\ \ga_9,\ x_4,\ x_8,\ x_9,\ z_4,\ z_6,\ z_{14}$    disappear from the action. As a result, the couplings with the above coefficients represent  only the cyclic symmetry and the Bianch identities.    So we ignore such terms in the DBI part.  Finally, we find that the couplings with coefficients   $\ga_4,\ t_2,\ x_2,\ x_3,\ z_{22},\ z_{29},\ z_{30}$ are total derivative terms, so they can be eliminated from the DBI part too. The final result for the DBI part  has no unknown coefficients! The couplings are those that appear in \reef{RTN1} and \reef{DBI2}.

\section{CS couplings}

In this section, using the mathematica package ``xAct'' \cite{CS}, we are going to write all couplings of one closed string RR state and two open NS strings   with unknown coefficients. We then constrain the coefficients by imposing the consistency of the couplings with the linear dualities and with the corresponding S-matrix element.  The S-matrix elements \reef{SCS} indicates that the world volume   couplings of D$_p$-brane in the CS part has  three parts. One is the couplings of  one $C_{p-3}$ and two gauge fields, another one is the couplings of one $C_{p-1}$, one gauge field   and one transverse scalar field, and the last one is the couplings of one  $C_{p+1}$ and two transverse scalar fields. Let us consider each case separately.

\subsection{One RR and  two gauge fields }

In this section we construct all possible couplings of one $C_{p-3}$  and  two gauge fields. Using the bulk equations of motion \reef{eom}, one finds there are   23 non-zero couplings, \ie  
\beqa
S_{caa}&\!\!\!\!\!=\!\!\!\!\!&\frac{\pi^2\alpha'^2T_{p}}{12}\int d^{p+1}x\,\epsilon^{a_0a_1\cdots a_{p}}\bigg[ \frac{1}{(p-2)!}\prt_{a}\cF^{(p-2)}_{a_3a_4\cdots a_p}\bigg(\frac{1}{2!}\ka_1\prt_{b}F_{a_1a_2}\prt^{b}F^{a}{}_{a_0}+\ka_2\prt_{a_0}F^{ba}\prt_{a_2}F_{ba_1}
\nonumber\\
&& \qquad\qquad\qquad\quad+\ka_3\prt_{a_2}F_{ba_1}\prt^{a}F^{b}{}_{a_0}+\ka_4\prt_{a_2}F_{ba_1}\prt^{b}F^{a}{}_{a_0}
+\frac{1}{2!}\ka_5\prt_{a_2}F_{b}{}^{a}\prt^{b}F_{a_0a_1}
\nonumber\\
&& \qquad\qquad\qquad\quad+\frac{1}{2!}\ka_6\prt^{a}F_{ba_2}\prt^{b}F_{a_0a_1}
+\ka_7\prt_{b}F^{b}{}_{a_0}\prt_{a_2}F^{a}{}_{a_1} +\frac{1}{2!}\ka_8\prt_{b}F^{b}{}_{a_0}\prt^{a}F_{a_1a_2}\bigg)\nonumber\\
&& \qquad\qquad\qquad\quad+\frac{1}{(p-3)!}\prt^{b}\cF^{(p-2)}{}_{ba_4\cdots a_p}\bigg(
\frac{1}{2!}\frac{1}{2!}\ka_{9}\prt_{a}F_{a_2a_3}\prt^{a}F_{a_0a_1}\nonumber\\
&& \qquad\qquad\qquad\quad+\ka_{10}\prt_{a_1}F^{a}{}_{a_0}\prt_{a_3}F_{aa_2} +\frac{1}{2!}\ka_{11}\prt^{a}F_{a_0a_1}\prt_{a_3}F_{aa_2}
\bigg)
\labell{caa}\\
&& \qquad\qquad\qquad\quad+\frac{1}{(p-3)!}\prt_{a}\cF^{(p-2)}_{ba_4\cdots a_p}\bigg(
\ka_{12}\prt_{a_1}F^{b}{}_{a_0}\prt_{a_3}F^{a}{}_{a_2}+\frac{1}{2!}\ka_{13}\prt_{a_3}F^{a}{}_{a_2}\prt^{b}F_{a_0a_1}
\nonumber\\
&& \qquad\qquad\qquad\quad+\frac{1}{2!}\ka_{14}\prt_{a_1}F^{b}{}_{a_0}\prt^{a}F_{a_2a_3} +\frac{1}{2!}\frac{1}{2!}\ka_{15}\prt^{b}F_{a_0a_1}\prt^{a}F_{a_2a_3}\bigg)
\nonumber\\
&& \qquad\qquad\qquad\quad+\frac{1}{(p-3)!}\prt_{a_4}\cF^{(p-2)}_{aa_3a_5\cdots a_p}\bigg(\frac{1}{2!}\ka_{16}\prt_{b}F_{a_1a_2}\prt^{b}F^{a}{}_{a_0}+\ka_{17}\prt_{a_2}F_{ba_1}\prt^{b}F^{a}{}_{a_0}
\nonumber\\
&& \qquad\qquad\qquad\quad+\ka_{18}\prt_{a_0}F^{ba}\prt_{a_2}F_{ba_1}
 +\ka_{19}\prt_{a_2}F_{ba_1}\prt^{a}F^{b}{}_{a_0}
+\ka_{20}\prt_{b}F^{b}{}_{a_0}\prt_{a_2}F^{a}{}_{a_1}
\nonumber\\
&& \qquad\qquad\qquad\quad+\frac{1}{2!}\ka_{21}\prt_{a_2}F_{b}{}^{a}\prt^{b}F_{a_0a_1} +\frac{1}{2!}\ka_{22}\prt^{a}F_{ba_2}\prt^{b}F_{a_0a_1}
+\frac{1}{2!}\ka_{23}\prt_{b}F^{b}{}_{a_0}\prt^{a}F_{a_1a_2}\bigg)\bigg]\nonumber
\eeqa
where  $\ka_i$ with $i=1,\cdots 23$ are the unknown constants that have to be found. In above equation, $\cF^{(p-2)}$ is the field strength of the RR potential $C_{p-3}$. One can easily verify that the above couplings are consistent with the T-duality transformations \reef{linear} when the killing index $y$ is a world volume index which is carried only by the RR field strength. When it is carried by the field strength of the gauge field, the consistency with T-duality requires the couplings of one $C_{p-1}$, one gauge field and one transverse scalar field which we consider them next.

\subsection{One RR, gauge field and one transverse scalar field}

All possible non-zero couplings of one RR potential $C_{p-1}$-form, one gauge field and one transverse scalar fields are the following:  
\beqa
S_{ca\chi}&\!\!\!\!\!=\!\!\!\!\!&\frac{\pi^2\alpha'^2T_{p}}{12}\int d^{p+1}x\,\epsilon^{a_0a_1\cdots a_{p}}\bigg[ \frac{1}{(p-1)!}\prt_{b}\cF^{(p)}_{ia_2a_3\cdots a_p}\bigg(\frac{1}{2!}\z_1\Om^{bai}\prt_{a}F_{a_0a_1}+\z_2\Om_{a_0}{}^{bi}\prt_{a}F^{a}{}_{a_1}\nonumber\\
&& \qquad\qquad\qquad\quad+\z_3\Om_{a_0}{}^{ai}\prt_{a}F^{b}{}_{a_1}+\z_4\Om^{bai}\prt_{a_1}F_{aa_0}+\z_5\Om_{a_0}{}^{ai}\prt_{a_1}F_{a}{}^{b}+\z_6\Om_{a_0}{}^{ai}\prt^{b}F_{aa_1}
\nonumber\\
&& \qquad\qquad\qquad\quad+\z_7\Om_{a}{}^{ai}\prt_{a_1}F^{b}{}_{a_0}+\frac{1}{2!}\z_8\Om_{a}{}^{ai}\prt^{b}F_{a_0a_1}\bigg)\nonumber\\
&& \qquad\qquad\qquad\quad+\frac{1}{(p-2)!}\prt^{b}\cF^{(p)}{}_{iba_3a_4\cdots a_p}\bigg(\frac{1}{2!}\z_9\Om_{a_0}{}^{ai}\prt_{a}F_{a_1a_2}+\z_{10}\Om_{a_0}{}^{ai}\prt_{a_2}{}F_{aa_1}\bigg)\nonumber\\
&& \qquad\qquad\qquad\quad+\frac{1}{(p-2)!}\prt_{b}\cF^{(p)}_{iaa_3a_4\cdots a_p}\bigg(\z_{11}\Om_{a_0}{}^{bi}\prt_{a_2}F^{a}{}_{a_1}
+\frac{1}{2!}\z_{12}\Om_{a_0}{}^{bi}\prt^{a}F_{a_1a_2}
\nonumber\\
&& \qquad\qquad\qquad\quad+\z_{13}\Om_{a_0}{}^{ai}\prt_{a_2}F^{b}{}_{a_1}
+\frac{1}{2!}\z_{14}\Om_{a_0}{}^{ai}\prt^{b}F_{a_1a_2}\bigg)\nonumber\\
&& \qquad\qquad\qquad\quad+\frac{1}{(p-3)!}\prt_{a_4}\cF^{(p)}_{iaba_3a_5\cdots a_p}\bigg(\z_{15}\Om_{a_0}{}^{ai}\prt_{a_2}F^{b}{}_{a_1}+\frac{1}{2!}\z_{16}\Om_{a_0}{}^{ai}\prt^{b}F_{a_1a_2}\bigg)\nonumber\\
&& \qquad\qquad\qquad\quad+\frac{1}{(p-2)!}\prt_{a_4}\cF^{(p)}_{iba_2a_3a_5\cdots a_p}\bigg(\z_{17}\Om_{a_0}{}^{bi}\prt_{a}F^{a}{}_{a_1}+\frac{1}{2!}\z_{18}\Om^{bai}\prt_{a}F_{a_0a_1}\nonumber\\
&& \qquad\qquad\qquad\quad+\z_{19}\Om^{bai}\prt_{a_1}F_{aa_0}
+\z_{20}\Om_{a}{}^{ai}\prt_{a_1}F^{b}{}_{a_0}\z_{21}\Om_{a_0}{}^{ai}\prt_{a}F^{b}{}_{a_1}+\z_{22}\Om_{a_0}{}^{ai}\prt_{a_1}F_{a}{}^{b}
\nonumber\\
&& \qquad\qquad\qquad\quad+\z_{23}\Om_{a_0}{}^{ai}\prt^{b}F_{aa_1}+\frac{1}{2!}\z_{24}\Om_{a}{}^{ai}\prt^{b}F_{a_0a_1}\bigg)\nonumber\\
&& \qquad\qquad\qquad\quad+\frac{1}{(p-1)!}\prt_{a_4}\cF^{(p)}_{ia_1a_2a_3a_5\cdots a_p}\bigg(\z_{25}\Om_{a}{}^{ai}\prt_{b}F^{b}{}_{a_0}+\z_{26}\Om^{bai}\prt_{b}F_{aa_0}
\nonumber\\
&& \qquad\qquad\qquad\quad+\z_{27}\Om_{a_0}{}^{bi}\prt_{a}F_{b}{}^{a}\bigg)\labell{cap}\\
&& \qquad\qquad\qquad\quad+\frac{1}{(p-2)!}\prt_{i}\cF^{(p)}_{aba_3a_4\cdots a_p}\bigg(\z_{28}\Om_{a_0}{}^{ai}\prt_{a_2}F^{b}{}_{a_1}+\z_{29}\Om_{a_0}{}^{ai}\prt^{b}F_{a_1a_2}\bigg)\nonumber\\
&& \qquad\qquad\qquad\quad+\frac{1}{(p-1)!}\prt_{i}\cF^{(p)}_{ba_2a_3a_4\cdots a_p}\bigg(\z_{30}\Om_{a_0}{}^{bi}\prt_{a}F^{a}{}_{a_1}+\z_{31}\Om^{bai}\prt_{a}F_{a_0a_1}
\nonumber\\
&& \qquad\qquad\qquad\quad+\z_{32}\Om^{bai}\prt_{a_1}F_{aa_0}+\z_{33}\Om_{a}{}^{ai}\prt_{a_1}F^{b}{}_{a_0}+\z_{34}\Om_{a_0}{}^{ai}\prt_{a}F^{b}{}_{a_1}
+\z_{35}\Om_{a_0}{}^{ai}\prt_{a_1}F_{a}{}^{b}\nonumber\\
&& \qquad\qquad\qquad\quad+\z_{36}\Om_{a_0}{}^{ai}\prt^{b}F_{aa_1}
+\frac{1}{2!}\z_{37}\Om_{a}{}^{ai}\prt^{b}F_{a_0a_1}\bigg)\nonumber\\
&& \qquad\qquad\qquad\quad+\frac{1}{p!}\prt_{i}\cF^{(p)}_{a_1a_2\cdots a_p}\bigg(\z_{38}\Om_{a}{}^{ai}\prt_{b}F^{b}{}_{a_0}
+\z_{39}\Om^{bai}\prt_{b}F_{aa_0}+\z_{40}\Om_{a_0}{}^{bi}\prt_{a}F_{b}{}^{a}\bigg)\bigg]\nonumber
\eeqa
where we have also imposed the bulk equations of motion \reef{eom}. In above equation  $\z_i$ with  $i=1,\cdots 40$ are the unknown constants that have to be found by consistency with dualities and with the S-matrix elements. 
One can easily verify that the above couplings are consistent with the T-duality transformations \reef{linear} when the killing index $y$ is a world volume index which is carried only by the RR field strength. This index can not be carried by the transverse scalar field. When it is carried by the field strength of the gauge field, the consistency with T-duality requires the couplings of one $C_{p+1}$ and two    transverse scalar fields which we consider them next.

\subsection{One RR and two transverse scalar fields}

All possible non-zero couplings of one RR potential $C_{p+1}$-form and two scalar fields after imposing the bulk equations of motion \reef{eom}   are the following:   
\beqa
S_{c\chi\chi}&\!\!\!\!\!=\!\!\!\!\!&\frac{\pi^2\alpha'^2T_{p}}{12}\int d^{p+1}x\,\epsilon^{a_0a_1\cdots a_{p}}\bigg[\frac{1}{(p+1)!}\prt_{b}\cF^{(p+2)}_{ca_0a_1\cdots a_p}\bigg(\rho_1\Om^{cai}\Om^{b}{}_{ai}+\rho_2\Om_{a}{}^{ai}\Om^{bc}{}_{i}\bigg) \nonumber\\
&& \qquad\qquad\qquad\quad-\frac{1}{(p+1)!}\prt^{b}\cF^{(p+2)}_{ba_0a_1\cdots a_p}\bigg(\rho_{3}\Om_{a}{}^{ai}\Om_{c}{}^{c}{}_{i}+\rho_{4}\Om_{cai}\Om^{cai}\bigg)\nonumber\\
&& \qquad\qquad\qquad\quad+\frac{1}{(p+1)!}\prt_{j}\cF^{(p+2)}_{ia_0a_1\cdots a_p}\bigg(\rho_{5}\Om_{a}{}^{ai}\Om_{c}{}^{cj} +\rho_{6}\Om_{ca}{}^{j}\Om^{cai}\bigg)\nonumber\\
&& \qquad\qquad\qquad\quad+\frac{1}{p!}\prt_{b}\cF^{(p+2)}_{ija_1\cdots a_p}\bigg(\rho_7\Om_{a_0}{}^{ai}\Om^{b}{}_{a}{}^{j}
+\rho_8 \Om_{a}{}^{ai}\Om^{b}{}_{a_0}{}^{j}\bigg)+\frac{\rho_9}{p!}\Om^{a}{}_{a_0}{}^{i}\Om^{cb}{}_{i}\prt_{c}\cF^{(p+2)}_{aba_1\cdots a_p}\nonumber\\
&& \qquad\qquad\qquad\quad+\frac{1}{p!}\prt^{c}\cF^{(p+2)}_{bca_1a_2\cdots a_p}\bigg(\rho_{10}\Om_{a_0}{}^{ai}\Om^{b}{}_{ai}+\rho_{11}\Om_{a}{}^{ai}\Om^{b}{}_{a_0i}\bigg)
\nonumber\\
&& \qquad\qquad\qquad\quad+\frac{1}{p!}\prt_{i}\cF^{(p+2)}_{jba_1a_2\cdots a_p}\bigg(\rho_{12}\Om_{a_0}{}^{ai}\Om^{b}{}_{a}{}^{j}+\rho_{13}\Om_{a}{}^{ai}\Om^{b}{}_{a_0}{}^{j}\bigg)
\labell{cpp}\\
&& \qquad\qquad\qquad\quad+\frac{1}{p!}\prt_{i}\cF^{(p+2)}_{jba_1a_2\cdots a_p}\bigg(\rho_{14}\Om_{a_0}{}^{aj}\Om^{b}{}_{a}{}^{i}+\rho_{15}\Om_{a}{}^{aj}\Om^{b}{}_{a_0}{}^{i}\bigg)\nonumber\\
&& \qquad\qquad\qquad\quad+\frac{1}{(p-1)!}\bigg(\rho_{16}\Om^{a}{}_{a_1}{}^{i}\Om^{b}{}_{a_0}{}^{j}\prt_{a}\cF^{(p+2)}_{ijba_2\cdots a_p}+\rho_{17}\Om^{a}{}_{a_1}{}^{j}\Om^{b}{}_{a_0}{}^{i}\prt_{j}\cF^{(p+2)}_{iaba_2\cdots a_p}\bigg)\nonumber\\
&& \qquad\qquad\qquad\quad+\frac{1}{(p-1)!}\prt_{a_4}\cF^{(p+2)}_{ijba_1a_2a_3a_5\cdots a_p}\bigg(\rho_{18}\Om_{a_0}{}^{ai}\Om^{b}{}_{a}{}^{j}+\rho_{19}\Om_{a}{}^{ai}\Om^{b}{}_{a_0}{}^{j}\bigg)\nonumber\\
&& \qquad\qquad\qquad\quad+\frac{\rho_{20}}{(p-1)!}\Om_{a_1}{}^{ai}\Om_{a_0a}{}^{j}\prt^{c}\cF^{(p+2)}_{ijca_2\cdots a_p}+\frac{\rho_{21}}{(p-1)!}\Om^{a}{}_{a_1i}\Om^{b}{}_{a_0}{}^{i}\prt^{c}\cF^{(p+2)}_{abca_2\cdots a_p}\bigg]\nonumber
\eeqa
where $\rho_i$ with $i=1,\cdots 21$ are the unknown constants.  One can easily verify that the above couplings are consistent with the T-duality transformations \reef{linear} when the killing index $y$ is a world volume index. So there is no D$_p$-brane coupling involving the RR potential $C_{p+3}$. The above couplings are also consistent with the S-duality of the D$_3$-brane action.
 
Now consider the sum of couplings \reef{caa}, \reef{cap} and \reef{cpp}, \ie
\beqa
S^{CS}_p&=&S_{caa}+S_{ca\chi}+S_{c\chi\chi}
\eeqa
They are not invariant under the linear T-duality transformations \reef{linear} for arbitrary coefficients. Imposing the invariance under T-duality \reef{Tconstraint}, one finds the following relations between the constants in the CS part: 
\beqa
&\!\!\!\!\!\!\!\!\!\!&\rho_2=-\rho_1,\ \rho_9=\rho_{10},\ \rho_{3}=-\frac{\rho_1}{2},\ \rho_{4}=\frac{\rho_1}{2},\ \rho_{11}=-\rho_{10},\ \rho_{15}=\z_{17}+\z_{18}-\z_{19}-\z_{20}\nonumber\\
&\!\!\!\!\!\!\!\!\!\!&+\z_{21}+\z_{23}+\z_{24}
-\z_{30}-\z_{31}+\z_{32}+\z_{33}
-\z_{34}-\z_{36}-\z_{37}-\rho_{12}-\rho_{13}-\rho_{14} \nonumber\\
&\!\!\!\!\!\!\!\!\!\!&  \z_8=-\z_1-\z_2-\z_{17}-\z_{18}+\z_{19}+\z_{20}-\z_{21}
-\z_{23}-\z_{24}-\z_3+\z_4-\z_6+\z_7,\ \nonumber\\
&\!\!\!\!\!\!\!\!\!\!&\ka_6=\z_9-\z_{10}-\ka_1+2\ka_{9}+2\ka_{10}-2\ka_{11}+\ka_{12}
-\ka_{13}-\ka_{14}+\ka_{15}+\ka_{16}-\ka_{17}+\ka_3
-\ka_{19}\nonumber\\
&\!\!\!\!\!\!\!\!\!\!&+\ka_4+\ka_{22}+\frac{1}{2}(\z_1-\z_{11}+\z_{12}-\z_{13}+\z_{14}+\z_{18}-\z_{19}
-\z_{21}-\z_{23}-\z_3-\z_4-\z_6),\ \nonumber\\
&\!\!\!\!\!\!\!\!\!\!&\rho_{17}=\frac{1}{2}(-\z_{11}+\z_{12}+\z_{13}-\z_{14}-\z_{18}+\z_{19}-\z_{21}-\z_{23}-2\z_{28}+2\z_{29}+\z_{31}-\z_{32}
\nonumber\\
&\!\!\!\!\!\!\!\!\!\!&+\z_{34}+\z_{36}+\rho_{12}+\rho_{14}),\ \rho_{19}=\frac{1}{2}(\z_{11}-\z_{12}+\z_{13}-\z_{14}-\z_{18}+\z_{19}+2\z_{20}\nonumber\\
&\!\!\!\!\!\!\!\!\!\!&-\z_{21}-\z_{23}-2\z_{24}+\z_{31}-\z_{32}-2\z_{33}+\z_{34}+\z_{36}+2\z_{37}
+\rho_{12}+2\rho_{13}+\rho_{14}+2\rho_{16}),\ \nonumber\\
&\!\!\!\!\!\!\!\!\!\!& \rho_{20}=\frac{1}{4}(-2\z_{10}-\z_{11}+\z_{12}-\z_{13}+\z_{14}+\z_{18}-\z_{19}-\z_{21}-\z_{23}-\z_{31}+\z_{32}+\z_{34}+\z_{36}\nonumber\\
&\!\!\!\!\!\!\!\!\!\!&+2\z_9
+\rho_{12}-\rho_{14}-2\rho_{16}-2\rho_{18}),\ \rho_5=\frac{1}{2}(\z_1-2\z_{25}+\z_3+\z_{31}-\z_{32}+\z_{34}+\z_{36}+2\z_{38}\nonumber\\
&\!\!\!\!\!\!\!\!\!\!&-\z_4+\z_6+\rho_{12}+\rho_{14}),\ \rho_6=\frac{1}{2}(-\z_1-2\z_{26}-\z_3-\z_{31}+\z_{32}-\z_{34}-\z_{36}+2\z_{39}+\z_4-\z_6\nonumber\\
&\!\!\!\!\!\!\!\!\!\!&-\rho_{12}-\rho_{14}),\ \rho_{7}=\frac{1}{2}(\z_1+\z_3-\z_{31}+\z_{32}+\z_{34}+\z_{36}+\z_4+\z_6+\rho_{12}-\rho_{14}),\ \rho_8=\frac{1}{2}(-\z1\nonumber\\
&\!\!\!\!\!\!\!\!\!\!&-2\z_{17}-2\z_{18}+2\z_{19}-2\z_2+2\z_{20}-2\z_{21}-2\z_{23}-2\z_{24}-\z_3+\z_{31}-\z_{32}-2\z_{33}+\z_{34}+\z_{36}\nonumber\\
&\!\!\!\!\!\!\!\!\!\!&+2\z_{37}+\z_4-\z_6+\rho_{12}+2\rho_{13}+\rho_{14}),\ \ka_8=\ka_{12}-\ka_{13}-\ka_{14}+\ka_{15}-\ka_{20}+\ka_{23}+\ka_7-\z_2\nonumber\\
&\!\!\!\!\!\!\!\!\!\!&-\z_{17}+\frac{1}{2}(-\z_1-\z_{11}+\z_{12}-\z_{13}+\z_{14}-\z_{18}
+\z_{19}-\z_{21}-\z_{23}-\z_3+\z_4-\z_6)\labell{TCS}
\eeqa
 The above constraints make the CS action to be consistent with the T-duality. There are still many constants that are not fixed yet.

Imposing the   constraints \reef{TCS}, one finds the couplings \reef{cap} are not consistent with S-duality for D$_3$-brane case. The S-duality requires, up to some total derivative terms, the couplings in \reef{cap} to be in the form of $\Om\prt (*\cF^T)\cM_0\prt^2\cB$. Using the expansion \reef{doub}, one finds the following relation between the constants in the CS part and the DBI part:  
\beqa
&\!\!\!\!\!\!\!\!\!\!&\z_4=1+\z_1+\z_{13}-\z_{14}+\z_2+\z_{17}-\z_{28}+\z_{29}+\z_{31}-\z_{32}\nonumber\\
&\!\!\!\!\!\!\!\!\!\!&\z_5=1-2\ga_1+\z_{13}-\z_{14}+\z_2+\z_{17}+\z_3+\z_{27}-\z_{28}+\z_{29}+\z_{34}-\z_{35}-\z_{40}\nonumber\\
&\!\!\!\!\!\!\!\!\!\!&\z_6=1-\z_{11}+\z_{12}-\z_2-\z_{17}-\z_{18}+\z_{19}-\z_{21}-\z_{23}-\z_3-\z_{28}+\z_{29}+\z_{31}-\z_{32}\nonumber\\
&\!\!\!\!\!\!\!\!\!\!&\z_{9}=\z_{10}+\z_{13}-\z_{14}-\z_{18}+\z_{19}-\z_{28}+\z_{29}+\z_{31}-\z_{32}\nonumber\\
&\!\!\!\!\!\!\!\!\!\!&\z_{30}=1-2\ga_1+\z_{13}-\z_{14}+\z_{17}-\z_{28}+\z_{29}\nonumber\\
&\!\!\!\!\!\!\!\!\!\!&\z_{36}=-2+2\ga_1+\z_{11}-\z_{12}-\z_{13}+\z_{14}+\z_{18}-\z_{19}+\z_{21}+\z_{23}+2\z_{28}-2\z_{29}-\z_{31}+\z_{32}-\z_{34}\nonumber\\
&\!\!\!\!\!\!\!\!\!\!&\z_{37}=1-\ga_3-\z_{11}+\z_{12}-\z_{20}
+\z_{24}-\z_{28}+\z_{29}+\z_{33}\nonumber\\
&\!\!\!\!\!\!\!\!\!\!&\z_{38}=1-2\ga_1+\z_{13}-\z_{14}+\z_2+\z_{17}+\z_{25}-\z_{28}+\z_{29}\nonumber\\
&\!\!\!\!\!\!\!\!\!\!&\z_{39}=-1+2\ga_1-\z_{13}+\z_{14}-\z_2-\z_{17}+\z_{26}+\z_{28}-\z_{29}\labell{SCS1}
\eeqa
 as well as the constraint \reef{last}. Imposing the above constraints, one finds not only the couplings \reef{cap}   but also the couplings \reef{caa} become consistent with the S-duality for D$_3$-brane, \ie the couplings $\prt\prt C_0\prt F\prt F$ in \reef{caa} and the couplings  $\prt\prt \Phi\prt F\prt F$ in the DBI part combine into the S-duality invariant structure  \reef{doub}. 

We now compare the couplings   with the S-matrix elements. Imposing the constraints \reef{TCS} and \reef{SCS1} into the action $S_p^{CS}$, one finds the resulting couplings are consistent with the  S-matrix elements \reef{SCS} provided that
\beqa
&&\rho_{14}=-\rho_{12},\ \ga_1=1\labell{smat}
\eeqa

The final  step is to ignore the couplings which are total derivative terms or the couplings which can be eliminated by the  Bianchi identities.    Imposing the constraints \reef{TCS}, \reef{SCS1} and \reef{smat} into the action, we find the terms with coefficient $\rho_{13}$  in \reef{cpp} are total derivative terms, so  $\rho_{13}$ can be eliminated from the physical couplings.   The terms with coefficients $\rho_{1}, \rho_{10}, \rho_{12}, \rho_{18}, \rho_{21}$   in \reef{cpp}   can be canceled by the Bianchi identity. When we write the field strengths  in \reef{cap} in terms of corresponding potentials, we find the  terms with coefficients   $\z_i$ with  $i=1, 3, 7, 10, 20, 23, 24, 25, 26, 33$ disappear, so these constants can be eliminated from \reef{cap} by the Bianchi identity.  Moreover, we find that terms with coefficients    $\z_i$ with $i=2, 11, 12, 13, 14, 15, 16, 17, 18, 19, 21, 22, 27,
28, 29, 31, 32, 34, 35, 40$ are total derivative terms. As a result, these terms can be ignored too. In the couplings \reef{caa}, the constants  $\ka_i$ with $i=7, 10, 11,  20, 23$ can be ignored by the Bianchi identities and the constants  $\ka_i$ with $i=1, 2, 3, 4, 5, 6, 12, 13, 14, 15, 16, 17, 18, 19, 21, 22$ can be ignored by total derivative terms.

 The final results  for the CS part are the couplings which appear in \reef{CS1}, \reef{CS2} and the following couplings:
 \beqa
S_{CS} &\supset&\frac{\pi^2\alpha'^2T_{p}}{12}\int d^{p+1}x\,\epsilon^{a_0a_1\cdots a_{p}}\bigg[\frac{\ga_3}{2!(p-1)!}  \Omega_{a}{}^{ai}\prt^{b}F_{a_1a_0}\prt_{i}\cF^{(p)}_{ba_2\cdots a_p}+\frac{\ga_3}{p!}\Om_{a}{}^{ai}\Om_{b}{}^{bj}\prt_{j}\cF^{(p+2)}_{ia_0\cdots a_p}\nonumber\\
&&\qquad\qquad- \frac{1-\ga_3}{(p-1)!}\Om_{a}{}^{ai}\Om_{a_0}{}^{bj}\prt_{a_3}\cF^{(p+2)}_{ijba_1a_2a_4\cdots a_p}+\frac{1-\ga_3}{p!}\Om_{a}{}^{ai}\Om_{a_0}{}^{bj}\prt_{b}\cF^{(p+2)}_{ija_1\cdots a_p} \bigg]\labell{final}
\eeqa
where we have also used the following identity in the second term:
\beqa
&&(p+1)\epsilon^{a_0a_1\cdots a_{p}} \Om_{a_0}{}^{bj}\prt_{j}\cF^{(p+2)}_{iba_1a_2\cdots a_p}=\epsilon^{a_0a_1\cdots a_{p}} \Om_{b}{}^{bj}\prt_{j}\cF^{(p+2)}_{ia_0a_1a_2\cdots a_p}
\eeqa
In proving the above identity we have used the totally antisymmetric property of the RR field strength  which can be used to replace the world volume index $b$ on the left-hand side by $a_0$. Using similar relation and writing the RR field strength in terms of RR potential, one can  prove the following identity:
\beqa
  -p \Om_{a}{}^{ai}\Om_{a_0}{}^{bj}\prt_{a_3}\cF^{(p+2)}_{ijba_1a_2a_4\cdots a_p}+ \Om_{a}{}^{ai}\Om_{a_0}{}^{bj} \prt_{b}\cF^{(p+2)}_{ija_1\cdots a_p}&=&0
\eeqa
  Using the above identity one finds that couplings in the second  line of \reef{final} are zero.
The   couplings in the first line of \reef{final} are consistent with the linear T-duality,  the S-duality and are zero when the scalar fields are on-shell. Note that the coupling in the first term  for  the case of D$_3$-brane can be written as    S-dual multiplet because $\Omega_{a}{}^{ai}\prt^{b}F^{cd}\prt_{i}H^{(3)}_{bcd}$ is zero by the Bianchi identity of the gauge field strength. Therefore, the coefficient  $\gamma_3$ can not be fixed by the linear dualities and by the S-matrix element of one closed and two open strings. It may be fixed by the open string pole of the S-matrix element of two closed strings and one open string at order $\alpha'^2$ or by the contact terms of the S-matrix element of three closed strings.  We expect the square of the second fundamental form appears in the world-volume curvatures as in \reef{RTRN}, \reef{r1} and \reef{r2}. The second fundamental forms   in the second term of \reef{final} can not be extended to the  curvature \reef{r2}, so we speculate  the coefficient of this term to be zero, \ie
\beqa
\gamma_3&=&0
\eeqa
It would be interesting to analyze in details the S-matrix element of two closed strings and one open string or the S-matrix element of three closed strings to confirm the above relation. 

Requiring the consistency of the D-brane effective action at order $\alpha'^2$ with S-matrix and with the linear dualities, we have found the couplings of one NSNS and two NS states in the DBI part to be \reef{RTN1} and \reef{DBI2}, and the  couplings of one RR and two NS states in the CS part to be \reef{CS1} and \reef{CS2}. On the other hand, the D-brane effective action at order $\alpha'^2$ should be invariant under supersymmetry and  $\kappa$ symmetry.   It would be interesting to verify the above couplings to be consistent with the supersymmetry and  $\kappa$ symmetry.

{\bf Acknowledgments}:   This work is supported by Ferdowsi University of Mashhad under grant 3/27085(1392/02/25).

\bibliographystyle{/Users/Nick/utphys} 
\bibliographystyle{utphys} \bibliography{hyperrefs-final}

\begin{thebibliography}{10}
 
 \bibitem{Bachas:1995kx}
  C.~Bachas,
  Phys.\ Lett.\  B {\bf 374}, 37 (1996)
  [arXiv:hep-th/9511043].
\bibitem{Douglas:1995bn}
  M.~R.~Douglas,
  arXiv:hep-th/9512077.

\bibitem{Bachas:1999um}
  C.~P.~Bachas, P.~Bain and M.~B.~Green,
  JHEP {\bf 9905}, 011 (1999)
  [arXiv:hep-th/9903210].
\bibitem{Garousi:1996ad}
  M.~R.~Garousi and R.~C.~Myers,
  Nucl.\ Phys.\  B {\bf 475}, 193 (1996)
  [arXiv:hep-th/9603194].
\bibitem{Hashimoto:1996kf}
  A.~Hashimoto and I.~R.~Klebanov,
  Phys.\ Lett.\  B {\bf 381}, 437 (1996)
  [arXiv:hep-th/9604065].
	
\bibitem{Green:1996dd}
  M.~B.~Green, J.~A.~Harvey and G.~W.~Moore,
  Class.\ Quant.\ Grav.\  {\bf 14}, 47 (1997)
  [arXiv:hep-th/9605033].
\bibitem{Cheung:1997az}
  Y.~K.~Cheung and Z.~Yin,
  Nucl.\ Phys.\  B {\bf 517}, 69 (1998)
  [arXiv:hep-th/9710206].
\bibitem{Minasian:1997mm}
  R.~Minasian and G.~W.~Moore,
  JHEP {\bf 9711}, 002 (1997)
  [arXiv:hep-th/9710230].
	
\bibitem{Garousi:2010ki} 
  M.~R.~Garousi,
  JHEP {\bf 1003}, 126 (2010)
  [arXiv:1002.0903 [hep-th]].
	
\bibitem{Garousi:2011fc} 
  M.~R.~Garousi,
  Phys.\ Lett.\ B {\bf 701}, 465 (2011)
  [arXiv:1103.3121 [hep-th]].
	
\bibitem{Garousi:2014oya} 
  M.~R.~Garousi,
  arXiv:1412.8131 [hep-th].
	
\bibitem{Robbins:2014ara} 
  D.~Robbins and Z.~Wang,
  JHEP {\bf 1405}, 072 (2014)
  [arXiv:1401.4180 [hep-th]].
	
\bibitem{Green:1997tv} 
  M.~B.~Green and M.~Gutperle,
  Nucl.\ Phys.\ B {\bf 498}, 195 (1997)
  [hep-th/9701093].
	
\bibitem{Liu:2013dna} 
  J.~T.~Liu and R.~Minasian,
  Nucl.\ Phys.\ B {\bf 874}, 413 (2013)
  [arXiv:1304.3137 [hep-th]].
	

\bibitem{TB}  
T. Buscher, Phys. Lett. B  {\bf 194} (1987) 59; B {\bf 201} (1988) 466.

\bibitem{Meessen:1998qm}
  P.~Meessen and T.~Ortin,
  Nucl.\ Phys.\  B {\bf 541}, 195 (1999)
  [arXiv:hep-th/9806120].
  
\bibitem{Bergshoeff:1995as}
  E.~Bergshoeff, C.~M.~Hull and T.~Ortin,
  Nucl.\ Phys.\  B {\bf 451}, 547 (1995)
  [arXiv:hep-th/9504081].
\bibitem{Bergshoeff:1996ui}
  E.~Bergshoeff, M.~de Roo, M.~B.~Green, G.~Papadopoulos and P.~K.~Townsend,
  Nucl.\ Phys.\  B {\bf 470}, 113 (1996)
  [arXiv:hep-th/9601150].

\bibitem{Hassan:1999bv}
  S.~F.~Hassan,
  Nucl.\ Phys.\  B {\bf 568}, 145 (2000)
  [arXiv:hep-th/9907152].
	
\bibitem{Garousi:2009dj} 
  M.~R.~Garousi,
  JHEP {\bf 1002}, 002 (2010)
  [arXiv:0911.0255 [hep-th]].
	
\bibitem{Gibbons:1995ap} 
  G.~W.~Gibbons and D.~A.~Rasheed,
  Phys.\ Lett.\ B {\bf 365}, 46 (1996)
  [hep-th/9509141].
	
\bibitem{Tseytlin:1996it}
  A.~A.~Tseytlin,
  Nucl.\ Phys.\ B {\bf 469}, 51 (1996)
  [hep-th/9602064].
	
\bibitem{Green:1996qg} 
  M.~B.~Green and M.~Gutperle,
  Phys.\ Lett.\ B {\bf 377}, 28 (1996)
  [hep-th/9602077].
	
\bibitem{Garousi:1998fg} 
  M.~R.~Garousi and R.~C.~Myers,
  Nucl.\ Phys.\ B {\bf 542}, 73 (1999)
  [hep-th/9809100].
	
	
\bibitem{CS}
T.~Nutma,
``xTras: a field-theory inspired xAct package for Mathematica,''
 	arXiv:1308.3493 [cs.SC].

  
  


\end{thebibliography}
\providecommand{\href}[2]{#2}\begingroup\raggedright
\endgroup

\end{document}